\documentclass[
  reprint,
  superscriptaddress,
  amsmath,
  amssymb,
  aps,
  prl
]{revtex4-2}

\usepackage{graphicx}

\begin{document}

\title{Explainable deep learning reveals the physical mechanisms behind the turbulent kinetic energy equation}

\author{Francisco Alcántara-Ávila}
\affiliation{Department of Aerospace Engineering, University of Michigan, Ann Arbor, MI 48109, USA}

\author{Andrés Cremades}

\author{Sergio Hoyas}
\affiliation{
Instituto Universitario de Matemática Pura y Aplicada, Universitat Politècnica de València, València 46022, Spain}

\author{Ricardo Vinuesa}
\email{rvinuesa@umich.edu.es}
\affiliation{Department of Aerospace Engineering, University of Michigan, Ann Arbor, MI 48109, USA}

\date{\today}

\begin{abstract}
In this work, we investigate the physical mechanisms governing turbulent kinetic energy transport using explainable deep learning (XDL). An XDL model based on SHapley Additive exPlanations (SHAP) is used to identify and percolate high-importance structures for the evolution of the turbulent kinetic energy budget terms of a turbulent channel flow at a friction Reynolds number of $Re_\tau = 125$. The results show that the important structures are predominantly located in the near-wall region and are more frequently associated with sweep-type events. In the viscous layer, the SHAP structures relevant for production and viscous diffusion are almost entirely contained within those relevant for dissipation, revealing a clear hierarchical organization of near-wall turbulence. In the outer layer, this hierarchical organization breaks down and only velocity-pressure-gradient correlation and turbulent transport SHAP structures remain, with a moderate spatial coincidence of approximately $60\%$. Finally, we show that none of the coherent structures classically studied in turbulence are capable of representing the mechanisms behind the various terms of the turbulent kinetic energy budget throughout the channel. These results reveal dissipation as the dominant organizing mechanism of near-wall turbulence, constraining production and viscous diffusion within a single structural hierarchy that breaks down in the outer layer.
\end{abstract}


\maketitle

The prediction and understanding of turbulent flows remain central challenges in modern fluid mechanics due to the highly nonlinear and multiscale nature of turbulence \cite{hul12,jim18,obe22}. Despite advances in high-fidelity numerical simulations, such as direct numerical simulation (DNS) \cite{kim87,lee15,hoy22}, the detailed analysis of the physical mechanisms driving the evolution of the velocity fields remains complex and computationally expensive. In this context, deep learning has emerged as a promising tool for capturing complex patterns in turbulence data and for generating predictive models that complement traditional approaches \cite{bru20,li_24}.

In particular, deep-learning models have demonstrated the ability to learn causal relationships when trained to predict the evolution of a flow \cite{cremades2025classically}. While previous studies have largely focused on interpreting predictions of the velocity-fluctuation field itself \cite{cremades2024}, analyzing the individual instantaneous terms of the turbulent kinetic energy (TKE) budget (production, dissipation, viscous diffusion, turbulent transport, and velocity-pressure-gradient correlation \cite{hoy08}) offers a more detailed and physically grounded perspective. Each term represents a distinct physical mechanism governing turbulence evolution, and their separate analysis can reveal insights that are not accessible from velocity fields alone.

To perform this analysis, we train a deep-learning model to predict the future evolution of each TKE budget term from the instantaneous velocity field. One way to interpret such predictions is to identify which regions of the velocity-fluctuation field are most relevant for the evolution of each physical mechanism. In this context, SHapley Additive exPlanations (SHAP) \cite{lundberg2017} provide a rigorous way to quantify the spatial importance of each of the grid points of the velocity-fluctuation field, enabling the identification of dominant flow structures. This approach has been shown to yield physically meaningful interpretations in previous studies of turbulent flows \cite{cremades2024}, supporting its use for this type of analysis. By linking these importance maps to the physical processes represented in the TKE budgets, SHAP enhances interpretability, offers deeper insight into flow physics, and provides a framework for using machine-learning-based predictions to extend the classical analysis of wall-bounded turbulence.

\begin{figure}
\includegraphics[width=\linewidth]{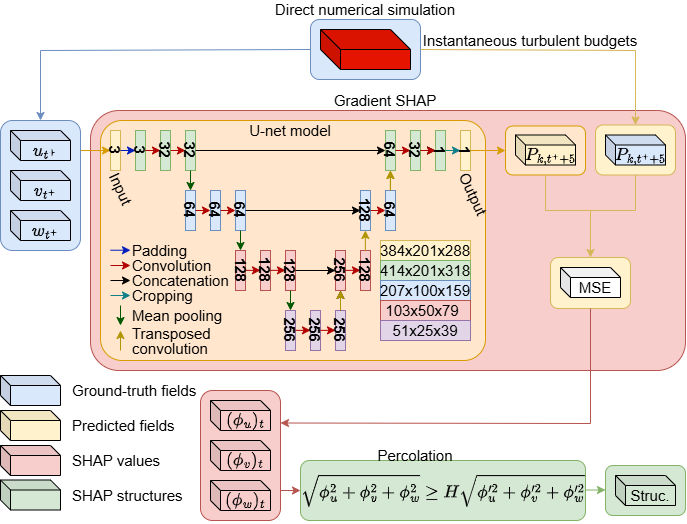}
\caption{Workflow diagram of the study, including data generation, neural-network training, SHAP analysis, and structure identification.}
\label{fig_methodology}
\end{figure}

The methodology followed in this work is represented in Figure~\ref{fig_methodology}. First, a direct numerical simulation (DNS) of a turbulent channel flow is conducted using the spectral code LISO, validated and used in a wide range of turbulence studies \cite{hoy22}. The LISO code simulates the continuity (\ref{continuity}) and Navier--Stokes equations (\ref{navier_stokes}) of an incompressible flow in a turbulent channel for a Newtonian fluid. These equations, using the Einstein summation convention, can be written as:
\begin{eqnarray}
\frac{\partial U_k}{\partial x_k} &&= 0, \label{continuity}\\
\frac{\partial U_i}{\partial t} + U_k\frac{\partial U_i}{\partial x_k} &&= -\frac{\partial P}{\partial x_i} + \nu \frac{\partial^2 U_i}{\partial x_k^2}, \label{navier_stokes}
\end{eqnarray}
\noindent where $U_i$ represents the instantaneous velocity for each of the three spatial directions (streamwise $x$, wall-normal $y$ and spanwise $z$), $P$ is the pressure, and $t$ is the time. The Reynolds decomposition can be defined as $U_i = \overline{U_i} + u_i$, where $\overline{U_i}$ is the average velocity in $x$, $z$ and $t$, and $u_i$ is the fluctuating velocity. In this work, the computational domain has dimensions $(L_x, L_y, L_z) = (8\pi h, 2h, 3\pi h)$ in the streamwise, wall-normal and spanwise directions, respectively. Here, $h$ is the half height of the channel. The flow is simulated at a friction Reynolds number of $Re_\tau = h u_\tau/\nu = 125$, high enough to obtain a fully turbulent flow \cite{pop00}. Note that $u_{\tau}=\sqrt{\tau_w}/\rho$ is the friction velocity, $\tau_w$ is the wall-shear stress, $\rho$ is the fluid density, and $\nu$ is its kinematic viscosity. The mesh used to discretize the domain is set to match the typical mesh resolution of other DNS works \cite{hoy22}. The data has been normalized in wall units, \textit{i.e.}, $y^+=y u_\tau/\nu$ and $u_i^+=u_i/u_\tau$.

Then, the instantaneous turbulent budget terms are calculated \cite{lum64}. The full derivation of the instantaneous turbulent budget terms is shown in End Matter. The definition of each term is given in the following equations: (\ref{production}) for the production $P_k$, (\ref{turbulent_transport}) for the turbulent transport $T_k$, (\ref{pressure}) for the velocity-pressure-gradient correlation term $\Pi_k$, (\ref{viscous_diffusion}) for the viscous diffusion $V_k$, (\ref{dissipation}) for the dissipation $\varepsilon_k$, (\ref{convective}) for the convective term $C_k$, and (\ref{instantaneous}) for an additional term $I_k$ which is present in the instantaneous form of the budget equations. Note that $\Pi_k$ term is used instead of the pressure-transport and pressure-strain splitting \cite{lee19} for simplicity of the analysis, and $u_1 = u$, $u_2 = v$, $u_3 = w$ are used for the rest of the discussion.

The deep-learning method employed here is the so-called U-net~\cite{ronneberger2015}, and it is represented in the orange box of Figure~\ref{fig_methodology}. We use a setup similar to that detailed in Cremades et al. \cite{cremades2025classically} using the same time interval, $\Delta t^+ = \Delta t u_\tau^2/\nu = 5$, for the prediction, but with the particularity here that the output has only one field, \textit{i.e.}, the corresponding term of the TKE budget (thus, we train one U-net for each term). The model is trained using a database of $10,000$ snapshots, reserving $20\%$ of them for validation. The training uses a RMSprop gradient-descent algorithm \cite{zou19} for minimizing the mean-squared error (MSE) of the predictions and it is trained until it ensures a relative error lower than $1\%$ for the training and validation datasets.

An additional set of $8000$ snapshots is used to calculate the importance of each grid point from the input field on the output, using a gradient-SHAP methodology~\cite{erion2021}. The gradient-SHAP algorithm is based on expected gradients~\cite{erion2021} an extension of the Shapley values~\cite{shapley1953} to a game with infinite players \cite{aumann2015values}.

The SHAP methodology identifies an importance score for each velocity-fluctuation component ($\phi_u$, $\phi_v$ and $\phi_w$) at each grid point, forming a three-dimensional importance field for each snapshot. This field is processed using percolation to identify coherent structures~\cite{Lozano2012}, comparing the local SHAP norm with the norm of the expected second-order momentum at the same wall-normal distance. The hyperbolic hole parameter $H$ is chosen to maximize the number of regions while minimizing the volume of the largest one \cite{del06}. Therefore, regions satisfying $\sqrt{\phi_u^2 + \phi_v^2 + \phi_w^2} > H\sqrt{\overline{\phi_u}^2 + \overline{\phi_v}^2 + \overline{\phi_w}^2}$ are referred to as SHAP structures. To further contextualize the SHAP-identified structures, we compare them with classical flow structures \cite{jeo95} commonly used in turbulence analysis, including intense Q events or Reynolds-stress structures \cite{Lozano2012}, streaks \cite{kli67}, and vortices \cite{cho90}, as well as the SHAP structures of the velocity-fluctuation field defined in Cremades et al.~\cite{cremades2025classically}.

While previous studies have focused on spectral analyses of the turbulent budget equations to investigate energy transfer across scales, notably the work of Lee and Moser \cite{lee19}, the present study adopts a complementary physical-space perspective, revealing where the flow structures responsible for each TKE budget term, documenting where they are located and how they overlap with other coherent structures across the wall-normal distance. Our work introduces a novel methodology to identify the most relevant flow structures, in the spirit of Jim\'enez \cite{jim18}, providing a physically grounded interpretation of the TKE budget that complements the spectral framework of Lee and Moser \cite{lee19}.

\begin{figure}
\includegraphics[width=\linewidth]{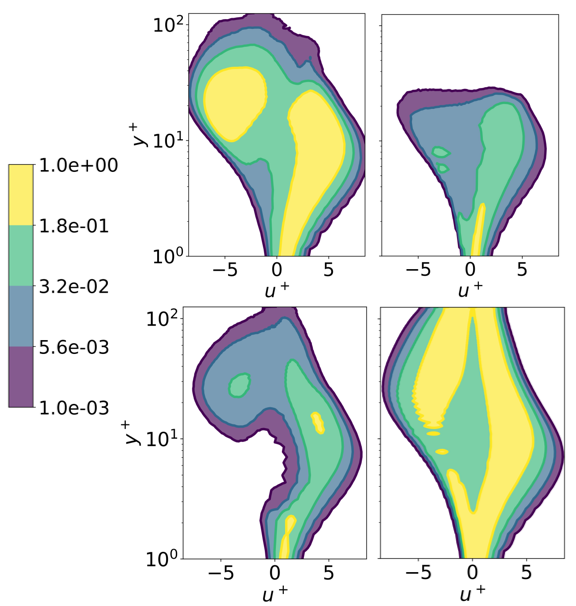}
\caption{Joint probability density functions (JPDFs) of the SHAP values based on the wall-normal distance and the streamwise velocity fluctuation for the production (top-left), dissipation (top-right), viscous diffusion (bottom-left) and turbulent transport (bottom-right).}
\label{fig_jpdfs}
\end{figure}

Figure~\ref{fig_jpdfs} shows the normalized joint probability density functions (JPDFs) of the SHAP structures for various TKE budget terms as a function of the wall-normal distance ($y^+$) and the streamwise velocity fluctuation ($u^+$). The first result is that most of these structures lie at $y^+ < 30$. Furthermore, in the buffer layer, around $y^+ \approx 15$, the density of these structures is even higher, especially for the production and viscous diffusion. In other words, the mechanisms associated with the TKE budget terms are strongly determined by the near-wall dynamics and the models correctly capture that most of the relevant activity occurs where the largest gradients and the most significant nonlinear interactions occur. Another important result is that very close to the wall, positive values of $u^+$ are more important than negative ones. In this context, the quadrant analysis proposed by Lu and Willmarth \cite{lu_73} is used to interpret the results. These positive $u^+$ values near the wall are related with sweep-type events (Q4), in contrast with the negative $u^+$ values farther away from the wall, which are related with ejection-type events (Q2). Therefore, sweep-like structures dominate over ejection-like ones when explaining the mechanisms for the TKE budget terms. This result highlights the relevance of the regions of high-speed flow approaching the wall, where the shear stress and the dissipation are stronger, a result which is consistent with the findings of previous works \cite{jim18,Lozano2012}.

With respect to the important structures to predict production (top-left) we observe that both sweep- and ejection-like structures are important. The maximum density is located at $y^+ \approx 15$, corresponding to the generation of high- and low-velocity streaks, which results in the maximum value of the averaged production term \cite{hoy08}. Regarding the structures for the dissipation, we observe that they completely vanish for $y^+ > 30$. This confirms that dissipation is dominated by small-scale motions associated with strong near-wall gradients, while the velocity fluctuations in the outer region provide no relevant information for predicting $\varepsilon_k$. Furthermore, the dissipation has a weak dependence on $u^+$. This is consistent with the nearly-isotropic nature of small-scale dissipation \cite{lee19}. The JPDF of the viscous diffusion structures exhibits a more complex and asymmetric distribution, with relatively important regions for $u^+ > 0$ and an S-shaped structure indicating a subtle dependence on $u^+$ and $y^+$. This suggests sensitivity to low-speed streaks, with a small wall-normal extent similar to the dissipation. Finally, the turbulent transport exhibits the largest wall-normal extent of important regions. This indicates that it connects near-wall dynamics with more external flow regions and requires information from a broad range of scales and wall-normal positions for accurate prediction. A strong bias towards $u^+>0$, particularly close to the wall, highlights the dominant role of sweep events in transporting turbulent energy and redistributing fluctuations towards the wall.

\begin{figure}
\includegraphics[width=\linewidth]{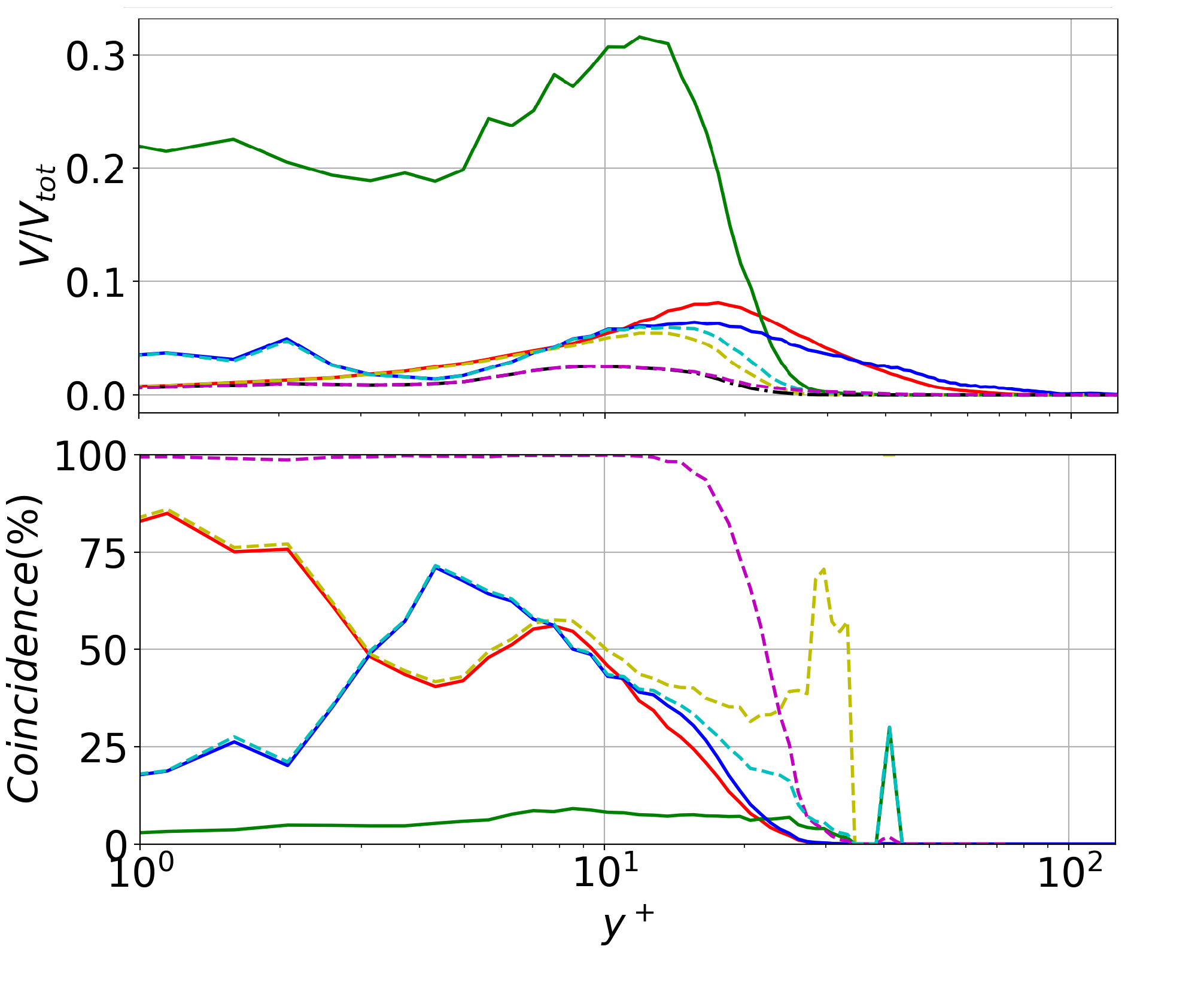}
\caption{(Top) Volume of SHAP structures of $P_k$ (solid red), $\varepsilon_k$ (solid green) and $V_k$ (solid blue). Intersection of SHAP structures of $P_k \cap \varepsilon_k$ (dashed yellow), $P_k \cap V_k$ (dashed magenta) and $\varepsilon_k \cap V_k$ (dashed cyan). Volume of the triple intersection $P_k\cap \varepsilon_k \cap V_k$ in dashed black. (Bottom) Percentage of coincidence of $P_k\cap \varepsilon_k \cap V_k$ normalized by the volume of SHAP structures of $P_k$, $\varepsilon_k$, $V_k$, $P_k \cap \varepsilon_k$, $\varepsilon_k \cap V_k$ and $\varepsilon_k \cap V_k$, with colors corresponding to the reference set shown in the top panel.}
\label{fig_nw_coincidences}
\end{figure}

When we analyze the data below $y^+ < 30$ we observe great similarities between the SHAP structures of dissipation, viscous diffusion, and production. This is represented in Figure~\ref{fig_nw_coincidences} as a function of $y^+$. In the top panel, we observe the fraction volume of the channel that contains a SHAP structure of each type: production, dissipation, or viscous diffusion, and their intersections. SHAP structures for dissipation are dominant in the near-wall region ($y^+<30$). On the other hand, production predominates over viscous diffusion close to $y^+\approx 15$, but closer to the wall the roles are inverted. When analyzing the intersections of two types of structures an important result is obtained: SHAP structures of production and viscous diffusion are a subset of the SHAP structures of dissipation. About $99 \%$ of production and viscous diffusion structures intersect with the dissipation structures below $y^+<10$, as can be observed in the lower panel of Figure \ref{fig_nw_coincidences}. As a conclusion the important structures to predict production and viscous diffusion are a subset of the important structures to predict dissipation. This result suggests that targeting or influencing the structures relevant for dissipation near the wall would inherently affect production and viscous diffusion as well. Therefore, dissipation structures act as the dominant phenomenon governing the other near-wall turbulent processes, highlighting them as key regions for potential flow control interventions.

\begin{figure}
\includegraphics[width=\linewidth]{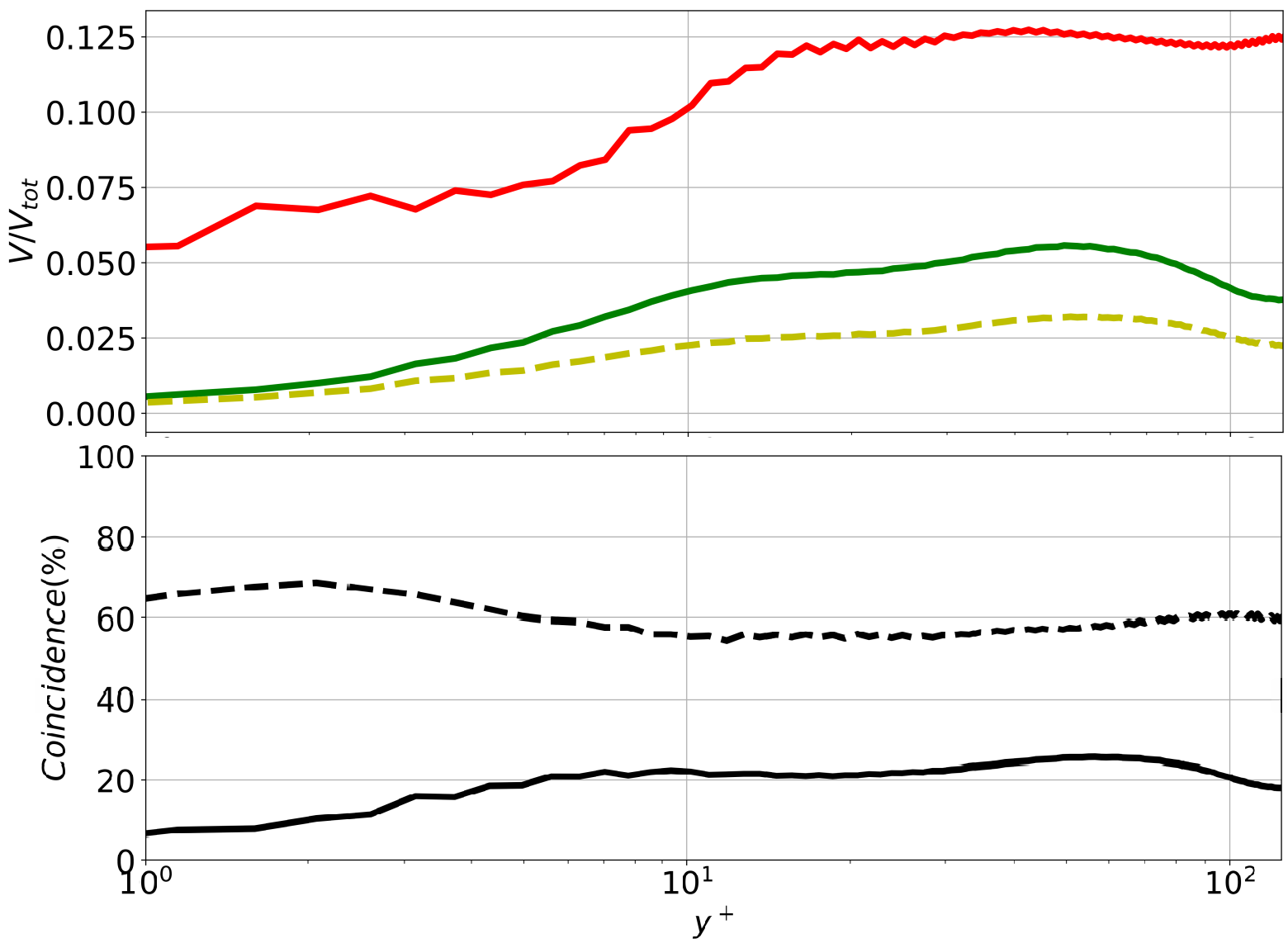}
\caption{(Top) volume of SHAP structures of $T_k$ (solid red), $P_k$ (solid green) and intersection of SHAP structures of $T_k \cap P_k$ (dasehd yellow). (Bottom) Coincidence percentage of the volume intersection of $T_k \cap P_k$ compared to the volume of SHAP structures of $T_k$ (solid line), $P_k$ (dashed line).}
\label{fig_ol_coincidences}
\end{figure}

While some similarities between the velocity-pressure-gradient correlation and turbulent transport SHAP structures exist in the near-wall region, they are less relevant than those observed for the previously analyzed terms. In contrast, in the outer layer of the flow, velocity-pressure-gradient correlation and turbulent transport SHAP structures are the only ones that persist. The reduced coincidence compared to the near-wall region highlights the more distributed and less tightly coupled nature of outer-layer turbulence, where multiple mechanisms contribute to energy redistribution across scales \cite{hoy08}. Away from the wall, however, these two terms exhibit the greatest coincidence, as shown in Figure~\ref{fig_ol_coincidences}. This coincidence indicates that a significant fraction of velocity-pressure-gradient correlation-related structures ($55$–$60$\%) is contained within turbulent-transport SHAP structures, suggesting a strong coupling between pressure effects and energy redistribution mechanisms in the outer layer. Furthermore, the nearly parallel evolution of the velocity-pressure-gradient correlation and turbulent transport SHAP structure volumes indicates that their spatial relevance is governed by the same wall-normal scaling and coherent structures.

\begin{figure}
\includegraphics[width=\linewidth]{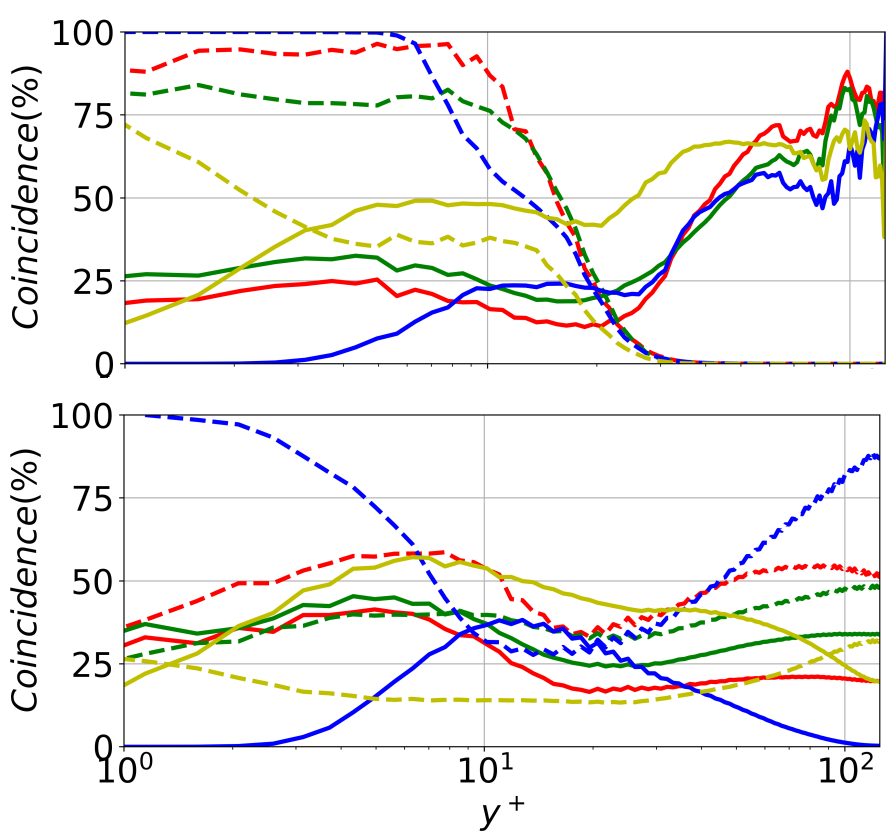}
\caption{Coincidence of classical structures with SHAP structures of $\varepsilon_k$ (top) and $T_k$ (bottom). Solid lines represents the percentage compared to SHAP structures of $\varepsilon_k$ (top) and $T_k$ (bottom). Dashed lines represent the percentage compared to the classical structures. Colors are for intersection with Q event structures (green), streaks (blue), vortices (yellow) and SHAP of velocity fluctuation structures (red).}
\label{fig_class_coincidences}
\end{figure}

Finally, we compare the SHAP-identified structures with classical turbulent flow structures. In this comparison, we focus on the dissipation and turbulent transport terms, since they are the dominant contributions in the near-wall and outer-layer regions, respectively. Figure~\ref{fig_class_coincidences} shows the corresponding coincidences for dissipation (top) and turbulent transport (bottom).

When comparing dissipation SHAP structures with classical structures, we find that Q events are largely contained within the dissipation ones (green dashed line). Streak structures are also embedded within the dissipation SHAP structures very close to the wall; however, this coincidence is not significant, as the number of streaks is nearly zero in this region ($y^+<3$). Farther from the wall, although the number of streaks increases, the percentage of coincidence rapidly decreases as the dissipation structures vanish. The strongest agreement is observed with the SHAP structures of the velocity fluctuations, for which approximately $90$\% are contained within the dissipation SHAP structures.

In the outer layer, none of the classical structures under study adequately represents the driving mechanisms of turbulent transport. Only Q events structures exhibit a moderate similarity, remaining below $50$\%. Furthermore, a local minimum at $y^+\approx15$ is observed for nearly all intersections, indicating that classical structures fail to capture the dominant mechanisms governing the turbulent budgets from the upper buffer layer to the channel center. Finally, we note that the similarity with vortices is below $50$\% in most of the relevant regions, a fact that makes them the least informative classical structures when describing the mechanisms behind the TKE budget terms. This result is in agreement with what  Cremades et al. 2025 \cite{cremades2025classically} documented for the velocity fluctuations. Similar conclusions hold for the remaining TKE budget terms, but are not shown here for brevity.

To summarize, SHAP provides a robust and interpretable framework to identify flow structures that are most relevant for predicting TKE budgets, bridging data-driven models and classical wall-turbulence physics. Based on SHAP values, we propose a new definition of ``important structures”, enabling quantitative comparisons across different  budget terms.

Near the wall ($y^+ < 10$), the SHAP structures relevant for production and viscous diffusion are largely a subset of dissipation SHAP structures, indicating that dissipation structures act as dominant ``envelope" regions. In the outer layer, the coincidence among structures is lower, with the largest overlap being between velocity-pressure-gradient correlation and turbulent-transport SHAP structures, reaching up to $60\%$, reflecting the more distributed and decorrelated nature of large-scale turbulence. This comparison further shows that classical structures barely align with SHAP-identified structures. Only the Q events represent important mechanisms for the dissipation near the wall, but none of them captures the important physics for $y^+\geq15$.

Overall, the SHAP framework reveals the hierarchical and spatial organization of turbulence, providing physically consistent insights that can inform both analysis and potential flow control strategies. In addition, these results demonstrate that SHAP is capable of extracting precise and meaningful information from extremely complex and highly nonlinear governing equations.

\begin{acknowledgments}
SH acknowledges support by CIN/AEI/10.13039/501100011033 and by ERDF, ``A way of making Europe”, under project PID2024-162480OB-I00.
\end{acknowledgments}

\bibliography{turbulence}

@article{bru20,
  title={Machine learning for fluid mechanics},
  author={Brunton, Steven L and Noack, Bernd R and Koumoutsakos, Petros},
  journal={Annual review of fluid mechanics},
  volume={52},
  number={1},
  pages={477--508},
  year={2020},
  publisher={Annual Reviews}
}

@article{cho90,
  title={A general classification of three-dimensional flow fields},
  author={Chong, Min S and Perry, Anthony E and Cantwell, Brian J},
  journal={Physics of Fluids A: Fluid Dynamics},
  volume={2},
  number={5},
  pages={765--777},
  year={1990},
  publisher={American Institute of Physics}
}

@article{del06,
  title={Self-similar vortex clusters in the turbulent logarithmic region},
  author={Del Alamo, Juan C and Jim{\'e}nez, Javier and Zandonade, Paulo and Moser, Robert D},
  journal={Journal of Fluid Mechanics},
  volume={561},
  pages={329--358},
  year={2006},
  publisher={Cambridge University Press}
}

@article{hoy08,
  title={{Reynolds number effects on the Reynolds-stress budgets in turbulent channels}},
  author={Hoyas, Sergio and Jim{\'e}nez, Javier},
  journal={Physics of Fluids},
  volume={20},
  number={10},
  year={2008},
  publisher={AIP Publishing}
}

@article{hoy22,
  title={Wall turbulence at high friction Reynolds numbers},
  author={Hoyas, Sergio and Oberlack, Martin and Alc{\'a}ntara-{\'A}vila, Francisco and Kraheberger, Stefanie V and Laux, Jonathan},
  journal={Physical Review Fluids},
  volume={7},
  number={1},
  pages={014602},
  year={2022},
  publisher={APS}
}

@article{hul12,
  title={Turbulent pipe flow at extreme Reynolds numbers},
  author={Hultmark, Marcus and Vallikivi, Margit and Bailey, Sean Christoper Collison and Smits, AJ},
  journal={Physical Review Letters},
  volume={108},
  number={9},
  pages={094501},
  year={2012},
  publisher={APS}
}

@article{jeo95,
  title={On the identification of a vortex},
  author={Jeong, Jinhee and Hussain, Fazle},
  journal={Journal of Fluid Mechanics},
  volume={285},
  pages={69--94},
  year={1995},
  publisher={Cambridge University Press}
}

@article{jim18,
  title={Coherent structures in wall-bounded turbulence},
  author={Jim{\'e}nez, Javier},
  journal={Journal of Fluid Mechanics},
  volume={842},
  pages={P1},
  year={2018},
  publisher={Cambridge University Press}
}

@article{kim87,
  title={Turbulence statistics in fully developed channel flow at low Reynolds number},
  author={Kim, John and Moin, Parviz and Moser, Robert},
  journal={Journal of Fluid Mechanics},
  volume={177},
  pages={133--166},
  year={1987},
  publisher={Cambridge University Press}
}

@article{kli67,
  title={The structure of turbulent boundary layers},
  author={Kline, Stephen J and Reynolds, William C and Schraub, Frederic Anthony and Runstadler, Peter W},
  journal={Journal of Fluid Mechanics},
  volume={30},
  number={4},
  pages={741--773},
  year={1967},
  publisher={Cambridge University Press}
}

@article{lee15,
  title={Direct numerical simulation of turbulent channel flow up to},
  author={Lee, Myoungkyu and Moser, Robert D},
  journal={Journal of fluid mechanics},
  volume={774},
  pages={395--415},
  year={2015},
  publisher={Cambridge University Press}
}

@article{lee19,
  title={Spectral analysis of the budget equation in turbulent channel flows at high Reynolds number},
  author={Lee, Myoungkyu and Moser, Robert D},
  journal={Journal of Fluid Mechanics},
  volume={860},
  pages={886--938},
  year={2019},
  publisher={Cambridge University Press}
}

@article{li_24,
  title={{Synthetic Lagrangian turbulence by generative diffusion models}},
  author={Li, Tianyi and Biferale, Luca and Bonaccorso, Fabio and Scarpolini, Martino Andrea and Buzzicotti, Michele},
  journal={Nature Machine Intelligence},
  volume={6},
  number={4},
  pages={393--403},
  year={2024},
  publisher={Nature Publishing Group UK London}
}

@article{lu_73,
  title={Measurements of the structure of the Reynolds stress in a turbulent boundary layer},
  author={Lu, SS and Willmarth, WW},
  journal={Journal of Fluid Mechanics},
  volume={60},
  number={3},
  pages={481--511},
  year={1973},
  publisher={Cambridge University Press}
}

@article{lum64,
  title={Spectral energy budget in wall turbulence},
  author={Lumley, JL},
  journal={The Physics of Fluids},
  volume={7},
  number={2},
  pages={190--196},
  year={1964},
  publisher={AIP Publishing}
}

@article{obe22,
  title={{Turbulence statistics of arbitrary moments of wall-bounded shear flows: A symmetry approach}},
  author={Oberlack, Martin and Hoyas, Sergio and Kraheberger, Stefanie V and Alc{\'a}ntara-{\'A}vila, Francisco and Laux, Jonathan},
  journal={Physical Review Letters},
  volume={128},
  number={2},
  pages={024502},
  year={2022},
  publisher={APS}
}

@Book{pop00,
  author	= {Pope, S. B.},
  isbn		= {9780521598866},
  lccn		= {99044583},
  publisher	= {Cambridge University Press},
  title		= {Turbulent Flows},
  year		= {2000}
}

@inproceedings{zou19,
  title={A sufficient condition for convergences of adam and rmsprop},
  author={Zou, Fangyu and Shen, Li and Jie, Zequn and Zhang, Weizhong and Liu, Wei},
  booktitle={Proceedings of the IEEE/CVF Conference on computer vision and pattern recognition},
  pages={11127--11135},
  year={2019}
}

@inproceedings{ronneberger2015,
  title={{U-net: Convolutional networks for biomedical image segmentation}},
  author={Ronneberger, Olaf and Fischer, Philipp and Brox, Thomas},
  booktitle={Medical Image Computing and Computer-Assisted Intervention--MICCAI 2015: 18th International Conference, Munich, Germany, October 5-9, 2015, Proceedings, Part III 18},
  pages={234--241},
  year={2015},
  organization={Springer}
}

@article{cremades2024,
  title={{Identifying regions of importance in wall-bounded turbulence through explainable deep learning}},
  author={Cremades, Andr{\'e}s and Hoyas, Sergio and Deshpande, Rahul and Quintero, Pedro and Lellep, Martin and Lee, Will Junghoon and Monty, Jason P and Hutchins, Nicholas and Linkmann, Moritz and Marusic, Ivan and others},
  journal={Nature Communications},
  volume={15},
  number={1},
  pages={3864},
  year={2024},
  publisher={Nature Publishing Group UK London}
}

@article{cremades2025classically,
  title={{Classically studied coherent structures only paint a partial picture of wall-bounded turbulence}},
  author={Cremades, Andr{\'e}s and Hoyas, Sergio and Vinuesa, Ricardo},
  journal={Nature Communications},
  volume={16},
  number={1},
  pages={10189},
  year={2025},
  publisher={Nature Publishing Group UK London}
}

@article{shapley1953,
  title={{A value for n-person games}},
  author={Shapley, Lloyd S and others},
  year={1953},
  journal={Princeton University Press Princeton}
}

@book{aumann2015values,
  title={Values of non-atomic games},
  author={Aumann, Robert J and Shapley, Lloyd S},
  year={2015},
  publisher={Princeton University Press}
}

@article{erion2021,
  title={{Improving performance of deep learning models with axiomatic attribution priors and expected gradients}},
  author={Erion, Gabriel and Janizek, Joseph D and Sturmfels, Pascal and Lundberg, Scott M and Lee, Su-In},
  journal={Nature Machine Intelligence},
  volume={3},
  number={7},
  pages={620--631},
  year={2021},
  publisher={Nature Publishing Group UK London}
}

@article{lundberg2017,
  title={{A unified approach to interpreting model predictions}},
  author={Lundberg, Scott M and Lee, Su-In},
  journal={Advances in Neural Information Processing Systems},
  volume={30},
  year={2017}
}

@Article{Lozano2012,
author = {Adrián Lozano-Durán and Oscar Flores and Javier Jiménez},
title = {{The three-dimensional structure of momentum transfer in turbulent channels}},
journal = {Journal of Fluid Mechanics},
year = {2012},
volume = {694},
pages={100-130}
}

\appendix
\section{Instantaneous turbulent budgets}

Applying the Reynolds decomposition to the Navier--Stokes equations (\ref{navier_stokes}) yields the expanded form of the instantaneous Navier--Stokes equations:
\begin{eqnarray}
\frac{\partial \overline{U_i}}{\partial t} + \frac{\partial u_i}{\partial t} +& \overline{U_k}\frac{\partial \overline{U_i}}{\partial x_k} + \overline{U_k}\frac{\partial u_i}{\partial x_k} + u_k\frac{\partial \overline{U_i}}{\partial x_k} + u_k\frac{\partial u_i}{\partial x_k} =\nonumber\\
=&-\frac{\partial \overline{P}}{\partial x_i} - \frac{\partial p}{\partial x_i} + \nu \frac{\partial^2 \overline{U_i}}{\partial x_k^2} + \nu \frac{\partial^2 u_i}{\partial x_k^2}. \label{navier_stokes_intantaneous}
\end{eqnarray}
The averaged Navier--Stokes equations are obtained by averaging Eq.~(\ref{navier_stokes_intantaneous}):
\begin{equation}
\frac{\partial \overline{U_i}}{\partial t} + \overline{U_k}\frac{\partial \overline{U_i}}{\partial x_k} + \frac{\partial \overline{u_iu_k}}{\partial x_k} = -\frac{\partial \overline{P}}{\partial x_i} + \nu \frac{\partial^2 \overline{U_i}}{\partial x_k^2}. \label{navier_stokes_averaged}
\end{equation}
The fluctuating Navier--Stokes equations (with subindex $i$) are obtained by subtracting (\ref{navier_stokes_intantaneous})$-$(\ref{navier_stokes_averaged}):
\begin{equation}
\frac{\partial u_i}{\partial t} + \overline{U_k}\frac{\partial u_i}{\partial x_k} + u_k\frac{\partial \overline{U_i}}{\partial x_k} + \frac{\partial u_iu_k}{\partial x_k} - \frac{\partial \overline{u_iu_k}}{\partial x_k} = - \frac{\partial p}{\partial x_i} + \nu \frac{\partial^2 u_i}{\partial x_k^2}. \label{navier_stokes_fluctuating}
\end{equation}
The instantaneous form of the turbulent budgets is obtained by computing the following operation: (\ref{navier_stokes_fluctuating})$_i u_j+$(\ref{navier_stokes_fluctuating})$_j u_i$. Using the continuity equation (\ref{continuity}) to simplify the resulting expression, the instantaneous budget equation for $u_i u_j$ can be written as:
\begin{eqnarray}
\frac{\partial u_i u_j}{\partial t} = \overbrace{- u_ju_k\frac{\partial \overline{U_i}}{\partial x_k} - u_iu_k\frac{\partial \overline{U_j}}{\partial x_k}}^{P_{ij}} \overbrace{- \frac{\partial u_i u_j u_k}{\partial x_k}}^{T_{ij}} \nonumber \\
+ \overbrace{p\left(\frac{\partial u_i}{\partial x_j} + \frac{\partial u_j}{\partial x_i}\right)}^{\Pi^s_{ij}} \overbrace{- \left(\frac{\partial u_i p}{\partial x_j} + \frac{\partial u_j p}{\partial x_i}\right)}^{\Pi^d_{ij}} + \nonumber\\
+ \overbrace{\nu\left(\frac{\partial^2 u_i u_j}{\partial x_k^2}\right)}^{V_{ij}} \overbrace{- 2\nu\left(\frac{\partial u_i}{\partial x_k}\frac{\partial u_j}{\partial x_k}\right)}^{\varepsilon_{ij}}\nonumber \\
\overbrace{-\overline{U_k}\frac{\partial u_i u_j}{\partial x_k}}^{C_{ij}} + \overbrace{u_j\frac{\partial \overline{u_iu_k}}{\partial x_k} + u_i\frac{\partial \overline{u_ju_k}}{\partial x_k}}^{I_{ij}}, \label{budget_final}
\end{eqnarray}
\noindent where the terms in the right-hand-side are referred as production, turbulent transport, pressure-strain, pressure-diffusion, viscous diffusion, convective and instantaneous. Note that the last two terms are $0$ for the averaged turbulent budget equation in a channel flow but they should be considered in this analysis. The equation for the instantaneous turbulent budget terms of the kinetic energy is obtained using:
\begin{equation}
k = \frac{1}{2}u_i u_i,
\end{equation}
\noindent and the different terms are defined as:
\begin{eqnarray}
P_k &=& -uv\frac{\partial \overline{U}}{\partial y}, \label{production}\\
T_k &=& - \frac{1}{2}\left(\frac{\partial uuu}{\partial x} + \frac{\partial uuv}{\partial y} + \frac{\partial uuw}{\partial z} + \frac{\partial uvv}{\partial x} + \frac{\partial vvv}{\partial y}\right.\nonumber\\
        &&\hspace{0.5cm}\left. + \frac{\partial vvw}{\partial z}+ \frac{\partial uww}{\partial x} + \frac{\partial vww}{\partial y} + \frac{\partial www}{\partial z}\right)
           \label{turbulent_transport}\\
\Pi_k &=& p\frac{\partial u}{\partial x} - \frac{\partial u p}{\partial x} + p\frac{\partial v}{\partial y}
           - \frac{\partial v p}{\partial y} + p\frac{\partial w}{\partial z} - \frac{\partial w p}{\partial z},
           \label{pressure}\\
V_k &=& \frac{\nu}{2}\left(\frac{\partial^2 uu}{\partial x^2} + \frac{\partial^2 uu}{\partial y^2} + \frac{\partial^2 uu}{\partial z^2}
           + \frac{\partial^2 vv}{\partial x^2} + \frac{\partial^2 vv}{\partial y^2}\right.\nonumber\\
           &&\hspace{0.3cm}\left.+ \frac{\partial^2 vv}{\partial z^2} + \frac{\partial^2 ww}{\partial x^2} + \frac{\partial^2 ww}{\partial y^2} + \frac{\partial^2 ww}{\partial z^2}\right),
           \label{viscous_diffusion}\\
\varepsilon_k &=& -\nu\left(\left(\frac{\partial u}{\partial x}\right)^2 + \left(\frac{\partial u}{\partial y}\right)^2 + \left(\frac{\partial u}{\partial z}\right)^2 + \left(\frac{\partial v}{\partial x}\right)^2 + \left(\frac{\partial v}{\partial y}\right)^2\right.\nonumber\\
&&\hspace{0.5cm}\left. + \left(\frac{\partial v}{\partial z}\right)^2 + \left(\frac{\partial w}{\partial x}\right)^2 + \left(\frac{\partial w}{\partial y}\right)^2 + \left(\frac{\partial w}{\partial z}\right)^2\right), \label{dissipation}\\
C_k &=& -\frac{\overline{U}}{2}\left(\frac{\partial uu}{\partial x} + \frac{\partial vv}{\partial x} + \frac{\partial ww}{\partial x}\right), \label{convective}\\
I_k &=& u\frac{\partial \overline{uv}}{\partial y} + v\frac{\partial \overline{vv}}{\partial y} + w\frac{\partial \overline{vw}}{\partial y},
           \label{instantaneous}
\end{eqnarray}
\noindent where the subscript $k$ refers to the kinetic energy.

\end{document}